\pdfoutput=1

\documentclass[journal]{IEEEtran}

\usepackage[utf8]{inputenc}
\usepackage{breqn}

\usepackage{algorithm}
\usepackage{algorithmic}
\usepackage{savesym}

\savesymbol{iint}
\usepackage{txfonts}
\restoresymbol{TXF}{iint}

\DeclareUnicodeCharacter{FF0C}{\textvisiblespace}

\ifCLASSINFOpdf
   \usepackage[pdftex]{graphicx}
   \graphicspath{{../pdf/}{../jpeg/}}
   \DeclareGraphicsExtensions{.pdf,.jpeg,.png}
\else
   \usepackage[dvips]{graphicx}
   \graphicspath{{../eps/}}
   \DeclareGraphicsExtensions{.jpg}
\fi

\usepackage[english]{babel}

\begin{document}

\title{Stealthy MTD Against Unsupervised Learning- based Blind FDI Attacks in Power Systems}

\author{
        Martin~Higgins,
        ~Fei~Teng,~\IEEEmembership{Member,~IEEE,}，
        and Thomas~Parisini,~\IEEEmembership{Fellow,~IEEE}

\thanks{The authors are with the Department of Electrical and Electronic Engineering, Imperial College London, London, SW7 2AZ, U.K. (Corresponding author: Dr Fei Teng, Email:  f.teng@imperial.ac.uk). T. Parisini is also affiliated with the Department of Engineering and Architecture, University of Trieste, Italy.}

\thanks{This work was partly supported by ESRC under Grant ES/T000112/1 and EPSRC Centre for Doctoral Training in Future Power Networks and Smart Grids (EP/L015471/1). Also, this work has been partially supported by the European Union's Horizon 2020 research and innovation programme under grant agreement No 739551 (KIOS CoE) and by the Italian Ministry for Research in the framework of the 2017 Program for Research Projects of National Interest (PRIN), Grant no. 2017YKXYXJ.
}

}

\markboth{IEEE Transactions on Information Forensics and Security, Submitted on 14 April~2020}%
{Shell \MakeLowercase{\textit{et al.}}: Bare Demo of IEEEtran.cls for IEEE Journals}

\maketitle

\begin{abstract}
This paper examines how moving target defenses (MTD) implemented in power systems can be countered by unsupervised learning-based false data injection (FDI) attack and how MTD can be combined with physical watermarking to enhance the system resilience. A novel intelligent attack, which incorporates  dimensionality reduction and density-based spatial clustering, is developed and shown to be effective in maintaining stealth in the presence of traditional MTD strategies. In resisting this new type of attack, a novel implementation of MTD combining with physical watermarking is proposed by adding Gaussian watermark into physical plant parameters to drive detection of traditional and intelligent FDI attacks, while remaining hidden to the attackers and limiting the impact on system operation and stability.

\end{abstract}


%
\IEEEpeerreviewmaketitle

\section{Introduction}

\IEEEPARstart{T}{he} modern power system is increasingly dependent on communication integrated devices for efficiency, reliability and control. The higher levels of inter-connectivity in the infrastructure and a ubiquitous use of communications have resulted in new types of vulnerabilities which have not been fully covered by the existing defense frameworks. Occurrences such as the 2015 cyber-attack against distribution companies in Ukraine \cite{Liang2017TheAttacks} have drawn attention to the field of defense against cyber-threats. The Ukraine attack took many months of infiltration and was successful in compromising the SCADA system and de-energizing a portion of the grid for a few hours. However, the attack itself was discovered almost instantly once implemented. If the attackers had opted for a stealthy attack type, such as FDI attacks, the attackers may have been able to continue attacking for months or years without being detected and the eventual consequences could have been much greater.   

FDI attacks, first outlined in \cite{Liu2011FalseGrids}, involve altering system measurements to corrupt a network operator's state estimation process and cause negative consequences such as line overloading or outage masking \cite {Liu2016MaskingAttacks}. A comprehensive review of FDI attacks can be found in \cite{Deng2017FalseSurvey}. FDI attacks need to remain undetected by the network operator to be effective. To this end, FDI attacks compete with bad data detectors (BDD) within state estimation processes. In modern energy management systems (EMS), the BDD at the power system level relies on weighted-least squares (WLS) and chi-squared error testing \cite{Liang2017ASystems}, meaning an attacker needs to structure the attack based on the system model in order to remain undetected. Initial models for FDI attacks assumed full knowledge of the system and full access to meter measurements within the system \cite{Liu2011FalseGrids}. An incomplete knowledge attack was introduced in \cite{Rahman2012FalseGrids}, which showed a system could be attacked with only partial knowledge of the system topology and a subset of meter measurements. In \cite{Esmalifalak2011StealthGrid} the blind FDI attack is introduced, which requires no system knowledge provided the attacker has access to all meters within the attacked grid system. The blind FDI attack uses independent component analyses (ICA) to map the inter-correlations of the visible meter measurements to create an approximation for the power flow model. 
A more effective version of the attack which utilizes partial susceptance knowledge was developed in \cite{Deng2019FalseGrid}, allowing an islanded approach where the visible or 'high knowledge' parts of the system could be attacked by the standard-FDI attack while low information areas by the blind approach. 
Some recent studies enhanced FDI attacks by combining with other forms of attacks, such as denial of service (DoS) attack \cite{Kurt2018Real-TimeGridb}.  

In addition, data-driven approaches have recently been applied to FDI attacks, although mostly from the defenders perspective \cite{Wang2017AGrids}\cite{Ahmed2019UnsupervisedForest}. 
In \cite{Wen2019AnGrids}, singular value decomposition is used to construct attack vectors without knowing the underlying system measurement matrix. 
In \cite{Kim2015SubspaceApproach}, two strategies using subspace separation are suggested: one aims to use estimated system subspace to hide attack vectors and another aims to mislead BDD so that non-attacked measurements are removed. These  methods  allow  for  admittance  values  to be estimated  but  require  a  large  number  of historical measurements. In \cite{Hao2015SparseGrids}, sparse FDI attacks against wide area measurement systems and defense methods are explored. Using historical data to mount FDI by using multiple linear regression model was outlined in \cite{Zhang2018CanSystems}. However, the above literature all focus on fixed network topology, while whether and how data-driven approaches can be applied to design FDI attacks under intentional or unintentional topology changes has not yet been investigated.

In fact, as FDI attacks are dependent on the characteristics of the physical system, 
a body of work has emerged to utilize the physical system to actively defend against the attacks. In particular, MTD is proposed through either transmission switching \cite{Wang2015EffectsNetworks} or admittance perturbation via distributed flexible AC transmissions (D-FACTS) devices \cite{Morrow2012TopologyInjection}
\cite{Liu2017ReactanceEstimation} to change physical system topology to proactively drive BDD. An analysis of MTD against FDI attacks is offered in \cite{Zhang2019AnalysisGrid} where they prove the susceptibility of isolated state measurements and design an algorithm for branch perturbation selection. Some limitations of MTD were explored in \cite{Li2019OnDevices}. With the increasing capability of the attackers, there are growing interests in the research community to design new forms of MTD which can hide its existence to the attacker. One of the key state-of-the-art papers in this field is \cite{Tian2019EnhancedGrids}, which presents an enhanced hidden MTD model to make the topology change invisible to an attacker via identifying alternative topology and state combinations under the same power flow profile. Whilst this method is clearly effective, it relies on being able to find alternative topology and states to maintain constant power flows, which can be computationally expensive and even infeasible in a system with limited acceptable state ranges. 








In this context, this paper examines the vulnerability of current MTD strategies under unsupervised learning-based FDI attacks and develops a new form of stealthy MTD to increase system resilience. Our main contributions are twofold: 

\begin{itemize}
        \item On the attacking front, this work introduces a novel new counter-MTD technique. Where previous FDI attacks have been designed against static systems, we seek to offer new attacking considerations in the presence of dynamic systems with MTD. 
        The proposed intelligent attack under zero system knowledge assumption combines  dimensionality reduction and unsupervised learning to identify underlying clusters associated with network topology and design the corresponding attack vector. The method is shown to be effective and stealthy against traditional MTD. 
        \item From the defensive perspective, we introduce a new implementation of MTD to drive detection against traditional and intelligent FDI attacks. The proposed defense strategy combines MTD and physical watermarking concept  \cite{Mo2015PhysicalOutputs}, for the first time, to add a Gaussian watermark into physical plant parameters. As the added watermark mimics the underlying noise of the system, the physical changes driven by MTD stay hidden. The physical watermarking is combined with cumulative error monitoring to spot minor but sustained changes in the system to trigger alarm. 
        




   
    %
        
      %
        
\end{itemize}

The rest of this paper is organized as follows. The problem formulation and underlying basis for FDI attacks and MTD through topology and parameter changes is outlined in section 2.  Section 3 details the design of the proposed intelligent attack, justification for algorithm selection and demonstration of its effectiveness in circumventing MTD. Section 4 proposes the  Gaussian style physical watermark in physical system parameters with cumulative error detection approach. Section 5 contains the results and analysis of the different types of MTD as applied to blind FDI attacks and Section 6 concludes the paper.

\section{Problem Formulation}
\subsection{State Estimation}

A static power system problem is considered, consisting of a set of $n$ state variables $\textbf{x} \in \mathbb{R}^{n\times1}$ estimated by analysing a set of $m$ meter measurements $\textbf{z} \in \mathbb{R}^{m\times1}$ and corresponding error vector $\textbf{e} \in \mathbb{R}^{m\times1}$ . The non-linear vector function $\textbf{h}(\textbf{.})$ relating meter measurements $\textbf{z}$ to states $\textbf{h}(\textbf{x}) = (h_1(\textbf{x}),h_2(\textbf{x}),...,h_m(\textbf{x}))^T$ is shown by 

\begin{equation}
    \textbf{z} = \textbf{h(x)} + \textbf{e}.
    \label{generalized state equation}
\end{equation}

With real power flow measurements under the non-linear expression defined by

\begin{equation}
    P_{ij} = V_{i}^2 g_{ij} - V_{i}V_{j}g_{ij}\cos{\Delta \theta_{ij}}-V_{i}V_{j}b_{ij}\sin{\Delta \theta_{ij}}.
   \label{Pac1}
\end{equation}



For simplicity and clarity, we first derive the initial formulation and condition based on the linear DC approximation of AC state estimation. A mathematical extension and simulations on original system are then preformed in later sections to demonstrate the applicability of the proposed methods in full AC state estimation. 

As a result, the matrix formulation, represented by a linear regression model as a function of the Jacobian $\textbf{H} \in \mathbb{R}^{m\times n}$ matrix and the state vector, can be expressed as: 





\begin{equation}
    \textbf{z} = \textbf{Hx} + \textbf{e}.
\end{equation}

The state estimation problem is to find the best fit estimate of $\hat{\textbf{x}}$ corresponding to the measured power flow values of $\textbf{z}$. Under the most widely used estimation approach, the state variables are determined by minimization of a WLS optimization problem as 

\begin{equation} \label{chisquaredfull}
 {\mathrm{min}}\,J(\textbf{x}) =  (\textbf{z}-\textbf{Hx})^T\textbf{W}(\textbf{z}-\textbf{Hx}).
\end{equation}

\textbf{W} is a diagonal $m \times m$ matrix consisting of the measurement weights.



A solution for a minimal $\textbf{J(x)}$ can be analytically obtained by taking the 1st derivative with respect to $\textbf{x}$ and solving for 0, yielding $\hat{\textbf{x}}$ defined by

\begin{equation} \label{estimate full}
    \hat{\textbf{x}} = (\textbf{H}^T\textbf{W}\textbf{H})^{-1}\textbf{H}^T\textbf{W}\textbf{z}.
\end{equation}

\subsection{Bad Data Detection}
The current approach in power systems operation for bad data detection is to use the 2-norm of the measurement residual with a detection threshold $\eta$ \cite{Monticelli1999StateApproach}. The residual $\textbf{r}$ is defined by the difference between the measured power flow values of $\textbf{z}$ and the value calculated from the estimated state values $\hat{\textbf{x}}$ and the known topology matrix $\textbf{H}$ 

\begin{equation} \label{residual}
    r =  ||\textbf{z} -\textbf{H}\hat{\textbf{x}}||_2.
\end{equation}

Assuming the errors of state variable $\textbf{x}$  are random, independent and follow a normal distribution with mean zero and unit $\mathcal{N}(0,\sigma^2)$, a chi-squared distribution model $\chi^{2}_{m-n,\alpha}$ with $m-n$ degrees of freedom and confidence interval $\alpha$  (typically 0.95 or 0.99) can be used to define the detection threshold as
\begin{equation}
    \eta = \sigma \sqrt{\chi^{2}_{m-n,\alpha}}.      
\end{equation}

If $r_t>\eta$ BDD alarms will trigger and the system operator will discard the result, removing the elements from the residual calculation with large values and replacing with an appropriate pseudo-measurement, based on historical data. 

\subsection{Constructing Attack Vectors}
In the case of an infinitely resourced and knowledgeable attack, the attacker can gain full access to the metering infrastructure and change measured power flows in any desired manner. 
In this case, it is trivial to design the attack to maintain a residual at a given value. The attacker can choose any linear combination of $\textbf{H}\textbf{c}$ where $\textbf{c} \in \mathbb{R}^{n\times1}$. The vector $\textbf{c}$ can be selected to have the desired impact on the state vector $\textbf{x}$:
\begin{equation}
    \textbf{z}_a = \textbf{z} + \textbf{a} =\textbf{z} + \textbf{H}\textbf{c}.
    \label{zattack}
\end{equation}


The 2-norm residual remains unchanged as shown below:

\begin{equation} \label{residualattack}
    r_a = \| (\textbf{z}+\textbf{a}) -\textbf{H}(\hat{\textbf{x}}+\textbf{c}) \|_2 = \| \textbf{z} -\textbf{H}\hat{\textbf{x}} \|_2.
\end{equation}



In a more realistic scenario, where the attacker has full access to the metering infrastructure but no understanding of how the network components interconnect or the branch admittance, the attacker has to commit a "blind" form of attack by estimating plausible attack vector models based on historical measurements. 
One way of achieving this is to utilize Blind Source Separation (BSS) techniques.  This scenario has been outlined previously in \cite{Esmalifalak2011StealthGrid}. \linebreak The relationship between the state variables in a power system and latent independent variables $\textbf{y}$ under a fixed topology $\textbf{H}$ can be described by

\begin{equation}
    \textbf{x} = \textit{f}(\textbf{H},\textbf{y}). 
    \label{zattack}
\end{equation}

In practice $\textbf{y}$ represents the loads of power system which vary independently while the topology is fixed but other underlying latent variables may exist for some systems. The state vector $x$ can be approximated as the first-order coefficient of the taylor expansion $\textbf{A}$ around $y$.  

\begin{equation}
    \textbf{x} \approx \textbf{A}\textbf{y}. 
    \label{zattack}
\end{equation}

Returning to the state estimation problem, the system states can then be expressed in terms of load such that
 
\begin{equation}
    \textbf{z} \approx\textbf{H}\textbf{A}\textbf{y} + \textbf{e}.
    \label{zattack}
\end{equation}

If the attacker can acquire $\textbf{H}\textbf{A}$, an attack vector can be constructed with a value selected for a change in power flows $\delta \textbf{y}$ shown by  

\begin{equation}
    \textbf{z}_{b} = \textbf{z}+\textbf{H}\textbf{A} \delta\textbf{y}.
    \label{zattack}
\end{equation}


A generalized form of blind source separation $\textbf{u}= \textbf{G}\textbf{v}$ can be used, where $\textbf{u}$ is the vector that can be directly observed,  $\textbf{G}$ is the fixed vector known as the mixing matrix and $\textbf{v}$ the underlying vector of signals. The state estimation can be constructed in an equivalent manner such that:

\begin{equation}
    \textbf{z} =\textbf{H}\textbf{A}\textbf{y} = \textbf{G}\textbf{y}.
    \label{zattack}
\end{equation}
Provided the errors follow a Gaussian distribution and do not contain gross errors, $\textbf{HA}$ can be extracted using independent component analysis as shown previously in \cite{Esmalifalak2011StealthGrid} \cite{Yu2015BlindGrid}.

\subsection{AC Extension of Blind Attack}


Similar to the DC attack, AC FDI attacks must satisfy the system model to remain hidden such that 

\begin{equation}
    \textbf{z}_a = \textbf{z}+ \textbf{a} = \textbf{h}(\textbf{x}+c).
    \label{Pac}
\end{equation}

This can be  done without topology information either using the geometric approach \cite{Chin2018BlindCommunications} or a historical measurement based replay approach. \textit{Chin} \textit{et al} showed that where the vector angle between the normal power flows and attacking vector was defined by 

\begin{equation}
     \textbf{z}^T\textbf{a} = cos(\psi)
    \label{zattack}
\end{equation}
the attack can bypass AC detection provided the vector space angle between the attacking vector and measurement vector was close to zero such that

\begin{equation}
     \textbf{z}^T\textbf{z}_a = 1.
    \label{zattack}
\end{equation}

Under these considerations, a regression model can be extracted to attack the system. Alternatively, in the case of limited information, the attacker can implement a replay style attack which reuses a previous vector from historical measurements such that

\begin{equation}
     \textbf{z}_t^a = \textbf{z}_{t-q}
    \label{zattack}
\end{equation}
where $q$ is used to denote a vector from a previous time period. Our AC simulations were built with this replay case in mind, but it should be noted both methods are susceptible to conventional MTD.

\subsection{MTD through Topology Changes}
Under AC state estimation, system measurements will consist of real power flows defined by (\ref{Pac1}) and reactive power by


\begin{dmath}
    Q_{ij} = -V_{i}^2(b_{ij}+b_{ij}^{sh}) + V_{i}V_{j}g_{ij}\cos{\Delta \theta_{ij}} \\ -V_{i}V_{j}b_{ij}\sin{\Delta \theta_{ij}}.
    \label{Qac}
\end{dmath}

For real power residual, error at the individual measurement level will be the difference between the measured flows and estimated value from the system model such that real power residual can be expressed as

\begin{dmath}
     r_{ij}^P = -P_{ij}^m +  V_{i}^2 g_{ij} - V_{i}V_{j} g_{ij}\cos{\Delta \theta_{ij}} \\ -V_{i}V_{j} b_{ij}\sin{\Delta \theta_{ij}}.
    \label{Pfac}
\end{dmath}
and reactive power flow residual can be expressed as 

\begin{dmath}
    r_{ij}^Q  = -Q_{ij}^m -V_{i}^2(b_{ij}+b_{ij}^{sh}) + V_{i}V_{j}g_{ij}\cos{\Delta \theta_{ij}} \\ -V_{i}V_{j}b_{ij}\sin{\Delta \theta_{ij}}.
    \label{Qac}
\end{dmath}

 
 In the AC state estimation, MTD can employ resistive as well as inductive components to introduce change. Alternatively, the SO can aim to force a state of non-convergence in the case of FDI which is done by violating the non-convergence criteria of the Newton-Raphson principle for power systems. Again, the alarm criteria will be the 2-norm value of the residual vector calculated by

\begin{equation}
    \textbf{r}_{ac} = \|\textbf{z} - \textbf{h}(\hat{\textbf{x}})\|_2.
    \label{generalized state equation}
\end{equation}

 We derive here the analytical expression of the impact on residual of topology change for a linear system under attack vector $\textbf{a}=\textbf{Hc}$. Using the WLS formulation,  $r_a$ can be expressed as 

\begin{equation} \label{residualhat}
     r_a = \| (\textbf{z}+\textbf{Hc}) -\textbf{H}(\textbf{H}^T\textbf{W}\textbf{H})^{-1}\textbf{H}^T\textbf{W}(\textbf{z}+\textbf{Hc}) \|_2.
\end{equation}

The attacker is assumed to have static topology knowledge and construct the injected attack vector $\textbf{z}_a$ as a function of the original topology $\textbf{H}_o$. The new topology with MTD applied is $\textbf{H}_n$, which is only known by the SO. As a result, the measurement vector under attack $\textbf{z}_a$ will be  

\begin{equation} \label{residualhat}
     \textbf{z}_a = \textbf{z}+\textbf{H}_{o}\textbf{c}.
\end{equation}

The SO estimates $\hat{\textbf{x}}$ via the WLS minimization using the visible $\textbf{z}_a$ and $\textbf{H}_n$. The min error estimate of $\hat{\textbf{x}}_n$ will utilize the new topology $\textbf{H}_n$ while the attack vector is developed based on the old topology $\textbf{H}_o$. Consequently, 
the new residual will be a product of the attack vector based on old topology $\textbf{H}_o\textbf{c}$ and the WLS estimation based on the new topology as

\begin{equation} \label{residualhat}
     r_n = \| \textbf{z}+\textbf{H}_{o}\textbf{c} - \textbf{H}_n(\textbf{H}^T_n\textbf{W}\textbf{H}_n)^{-1}\textbf{H}^T_n\textbf{W}(\textbf{z}+\textbf{H}_{o}\textbf{c})\|_2.
\end{equation}

Defining WLS minimization factor for the new topology as $\textbf{F}_n$, which is fixed for a given topology as ${\textbf{F}}_n = (\textbf{H}^T_n\textbf{W}\textbf{H}_n)^{-1}\textbf{H}^T_n\textbf{W}$, the residual 2-norm can be rewritten as

\begin{equation} \label{residualhat}
         r_n = \| \textbf{z}+\textbf{H}_{o}\textbf{c} - \textbf{H}_n\textbf{F}_n(\textbf{z}+\textbf{H}_{o}\textbf{c})\|_2.
\end{equation}

Considering the old topology $\textbf{H}_o$ as a function of the new and system change $\textbf{H}_n + \Delta\textbf{H}$, the residual in terms of the new topology can hence be calculated as


\begin{equation} \label{residualhat}
          r_n = \| \textbf{z}+(\textbf{H}_n + \Delta\textbf{H})\textbf{c} - \textbf{H}_n\textbf{F}_n(\textbf{z}+(\textbf{H}_n + \Delta\textbf{H})\textbf{c})\|_2.
\end{equation}

$\textbf{H}_n\textbf{F}_n\textbf{H}_n$ is the idempotent matrix of $\textbf{H}$ and therefore $\textbf{H}_n\textbf{F}_n\textbf{H}_n\textbf{c} = \textbf{H}_n\textbf{c}$  the expression can be rearranged into 



\begin{equation} \label{residualhat1}
          r_n = \|(1 - \textbf{H}_n\textbf{F}_n)\textbf{z}+(1-\textbf{H}_n\textbf{F}_n)\Delta\textbf{H}\textbf{c}\|_2.
\end{equation}







As shown in (\ref{residualhat1}), any $\Delta\textbf{H}$ will change the residual value $\textbf{r}_{n}$. The aim of defender is to select a value for $\Delta\textbf{H}$ such that under attack vector $\textbf{H}_{o}\textbf{c}$, the new residual exceeds the alarm criteria (usually chi-squared criteria)  $r_n > \sigma \sqrt{\chi^{2}_{m-n,\alpha}}$.

\section{Clustering to Circumvent MTD}

This section investigates how data-driven approach can be applied to explore the vulnerability of existing MTD. In particular, an efficient method is proposed to identify changes in the network caused by the implementation of D-FACTS or switching through analysing the resultant power flow profiles. By doing so, the attacker can ensure only data points corresponding to the current configuration are used to create the blind attack. The proposed attack flow follows:

\begin{enumerate}
    \item Observations of historical power flows are clustered into groups.
    \item The clustering algorithm identifies the current power flow set to find corresponding measurements for the attack model.
    \item The blind attack model is developed using only the data corresponding to the current power flow profile cluster.
\end{enumerate}

\begin{figure}[t]
\centering
\includegraphics[width=3.0in]{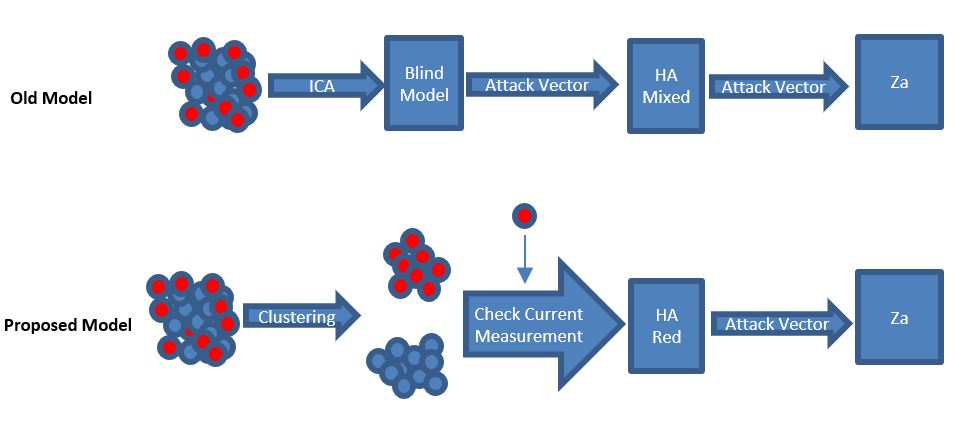}
\caption{Proposed algorithm process to circumvent MTD. Red and blue points corresponded to observed power flows from 2 different network configurations.}

\label{clusterprocess}

\end{figure}

A simple example of this process is illustrated and compared with the normal blind attack in Figure  \ref{clusterprocess}. To achieve this, we propose a combination of data prepossessing via T-distributed stochastic neighbour embedding (T-SNE) for dimensionality reduction followed by density based spatial clustering of application with noise (DBSCAN) to classify the data sets. We outline the justification for our chosen methods below.

\subsection{Attack Design Considerations}
 Power transmission systems are by their very nature large. To design such data-driven attacks, one of the key considerations is to maintain the feasibility of implementation in real-time operation of large scale systems. Therefore, it is essential to circumvent the curse of dimensionality (CoD) within the context of this attack. We hence explore the use of T-SNE to reduce the dimensionality of data sets before applying the clustering algorithm. In addition, due to the blind nature of the attack, no prior knowledge of the number of underlying topologies can be assumed and therefore an unsupervised learning method, DBSCAN in this case, is developed.

\subsubsection{T-SNE for Dimensionality Reduction}

T-SNE is a form of dimensionality reduction which works by constructing probability distributions over pairs of objects containing high dimensionality  \cite{VanDerMaaten2008VisualizingT-SNE}. T-SNE considers a set of $N$ high dimension objects. $d_i$ and $d_j$ are two points within this set. $\sigma_i$ is the variance of the Gaussian centred on data point $d_i$. The closeness of these data points is defined by the conditional probability $p_{j|i}$ that point $d_j$ would select $d_i$ as a neighbour given that the neighbours are picked proportionately to a Gaussian centred around $d_j$. This is given by

\begin{equation} \label{residualhat}
          p_{j|i} =\frac{\exp(-||d_i-d_j||^2/2\sigma_i^2)}{\sum_{k\neq i} \exp(-||d_i-d_j||^2/2\sigma_i^2)}.
\end{equation}

The aim of T-SNE is to reduce these points into their low dimensional counterparts $g_j$ and $g_i$. These have an equivalent conditional probability $q_{j|i}$ defined by 

\begin{equation} \label{dimensionreduc}
          q_{j|i} =\frac{\exp{(-||g_i-g_j||^2)}}{\sum_{k\neq i} \exp((-||g_i-g_j||^2))}.
\end{equation}

If the map points $g_j$ and $g_i$ correctly model the similarity between the high dimensional sets, the conditional probabilities $p_{j|i}$ and $q_{j|i}$ will be equal.  The positions of $g_i$ and $g_j$ are determined via  gradient descent between the distributions $p$ and $q$, and is used to minimize the Kullback-Leiber (KL) divergence via cost function $C$ \cite{ESilva2010MachineTesting} shown by

\begin{equation} \label{residualhat}
          C =\sum_i KL(P_i||Q_i)=\sum_i\sum_{j}p_{j|i}\log\frac{p_{j|i}}{q_{j|i}},
\end{equation}
where $P_i$ is the conditional probability distribution over all data points given data point $d_i$ and $Q_i$ is the conditional probability distribution over every other map point, given map point $g_i$.

Native T-SNE itself has a time complexity of $O(n^2)$ but this can be reduced to $O(n)$ by using optimization techniques as discussed in \cite{Pezzotti2017ApproximatedAnalytics}. The brunt of the computational load is therefore taken by T-SNE which reduces the measurements of the network power flows into 2-dimensional space. 

There are other possible unsupervised approaches for dimensionality reduction. Linear reduction algorithms such as principle component analysis (PCA) are one such example. PCA performs linear mapping to lower dimensional spaces and unlike T-SNE is deterministic rather than probabilistic. PCA being a linear algorithm means that it does have some computational benefits. However, PCA cannot represent complex polynomial relationships in the same way T-SNE can. Also, the KL divergence minimisation that T-SNE employs means much of the local structure of data is preserved in T-SNE which it is not to the same degree in PCA. We also consider that with the stated purpose of identifying like groupings of points T-SNE is also the most appropriate choice. The probabilistic neighbour assessment approach of T-SNE seeks to identify neighbours specifically which makes the output trend towards close and distinct cluster groups emerging (as shown in Figure \ref{gaussianvstp1}). This makes it easy to identify groups for the next section of the attack algorithm (clustering and model building) to operate over.     

\subsubsection{DBSCAN for Unsupervised Learning}

When the dimension reduced data set is received, a cluster algorithm will be applied to identify the underlining clusters of the data. Due to the blind nature of the attack, we propose using DBSCAN for the unsupervised clustering portion of this attack. The DBSCAN algorithm works as follows:

\begin{enumerate}
    \item An initial starting point is randomly selected. This point is then marked as visited.  
    \item The points adjacent to this point defined by $\epsilon$ are counted and added to a set
    \item If the number of points exceeds the defined min point value the initial point is defined as a new cluster. This process is continued for all points in the neighbourhood
    \item If the number of points is less than the min the point is defined as noise
    \item These steps are repeated until the whole set has been clustered. 
\end{enumerate}

DBSCAN  shows good benchmark performance against other forms of unsupervised learning \cite{McInnesBenchmarkingAlgorithms} and also offers several relevant advantages to this form of the attack.  DBSCAN has a time complexity of $O(n^2)$ but this can be reduced to $O(n \log n)$ with parameter optimisation \cite{Ester1996ANoise}, unlike hierarchical clustering which has a time complexity of $O(n^3)$ and is highly computationally intensive for large systems by comparison. DBSCAN also does not require pre-specification of the number of clusters (making it more appropriate for a blind style attack) and is robust against outlying data points and noise. Density-based Local Outlier Factor (LOF) was also considered and has previously been seen for FDI attack detection in \cite{Konstantinou2019AAttacks}. In many ways LOF is similar to DBSCAN but is more specialised for anomaly detection as opposed to direct clustering.

\subsection{Intelligent Blind FDI Attack}

This sub-section details the proposed intelligent blind FDI, as outlined in Algorithm \ref{algo1}. Once the attacker obtained adequate amount of measurement data, the attacker can initiate the attack algorithm. When the latest measurement data arrive, 
initially, T-SNE is applied for dimensionality reduction of the sets of power flow observations into a two dimensional space. The reduced form of the data set is then clustered via the DBSCAN algorithm into distinct subgroups of like measurements and the one corresponding to the current system topology is identified. The mixing matrix is subsequently derived based on this subgroup of data by using independent component analysis as per the normal blind attack. A vector of false data $\textbf{z}_a$ containing the desired attack bias will be then calculated based on the mixing matrix.

Ultimately, the attacker does not know which model corresponds to the base case or the case with MTD implemented. The attacker simply knows there are multiple distinct underlying models and creates a series of models equal to the number of clusters. The attacker may be able to guess based on how the topologies represent in terms of timing which is the base (no MTD case) but this is largely irrelevant for the attack. A minimum cluster size check will also be implemented to ensure the attack has sufficient data to create the blind model. 

 



\begin{algorithm}
\caption{DBSCAN Blind-ICA attack}

{\textbf{Input:} A set of power flow observations $\textbf{z}_{obvs}$}

 {\hspace{1cm}\textbf{Y} = tsne (\textbf{z}_{obvs}) \%Dimensionality reduction }

 {\hspace{1cm}\textbf{idx} = dbscan(Y,mpts,$\epsilon$) \% Cluster power flows}

 {\hspace{0.5cm}\textbf{For} i = 1:length(unique(idx))} \% Assign pf to cell

\hspace{1cm}j = [j,idx] \% Assign cluster to obvs

\hspace{1cm}\textbf{A}\{i\} = j(j(:,l)==i,:)\% Assign obvs to cell

 \hspace{1cm}{c = j(end)} \%check what profile current \textbf{z} is

\hspace{1cm} {\textbf{z} = A(c)} \% Select only corresponding \textbf{z} measurements

\hspace{1cm}{\textbf{HA} = FastICA(\textbf{z})} \% Run fastica for \textbf{HA}

\hspace{1cm}{   $\textbf{z}_a = \textbf{z}+\textbf{H}\textbf{A} \delta \textbf{y}$ } \% Apply attack vector

\textbf{Output:} false data $\textbf{z}_a$

\label{algo1}

\end{algorithm}

\subsection{Performance Analysis}
\begin{figure}[t]
\centering
\includegraphics[width=2.8in]{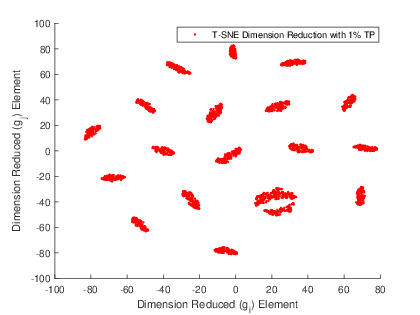}

\caption{Power flow profile observations of 1\% admittance perturbation MTD applied to 19 lines intermittently under T-SNE dimensionality reduction. The X and Y axis values are non-dimensional probabilistic reductions ($g_i$ \& $g_j$ from equation \ref{dimensionreduc}). The system is reduced from a 34 dimension meter IEEE 14-bus system.}
\label{gaussianvstp1}
\end{figure}

\begin{figure}[t]
\centering
\includegraphics[width=2.8in]{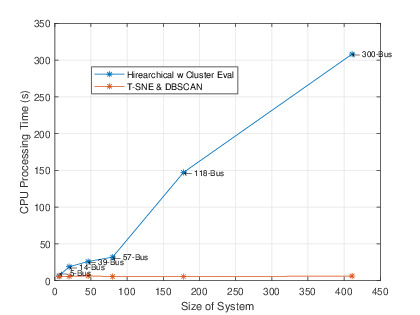}

\caption{CPU processing time for the combined T-SNE/DBSCAN algorithm with an equivalent hierarchical method with embedded cluster selection performed on systems of increasing size. Performed for 1000 observations.}
\label{computationresults}

\end{figure}

To demonstrate the performance of the proposed algorithm, a case study is carried out on a system with 14 lines equipped with D-FACTS for MTD. As shown in Figure \ref{gaussianvstp1}, the proposed algorithm successfully identify and cluster the potential topology sets, even only minor changes on topology (1\% of base admittance) are applied.   
The computational performance of T-SNE and DBSCAN for different IEEE standard systems is shown in Figure \ref{computationresults} and compared with hierarchical clustering with embedded cluster evaluation. The case studies are performed for 1000 sets of observations. We note that for small scale systems (such as the 5-bus case) the computational performance are similar but as the system becomes larger, the time to completion grows quickly for hierarchical clustering. 

\subsubsection{Real-Time Attacks in Large Systems}
\begin{figure}[t]
\centering
\includegraphics[width=2.8in]{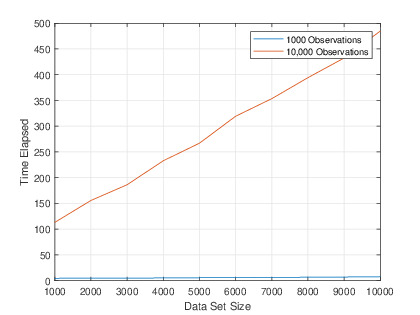}

\caption{CPU processing time for the combined T-SNE and DBSCAN algorithm for increasing size of random data array upto 10,000 data points. Performed for 1000 and 10,0000 observations.}
\label{largecomputationresults}

\end{figure}

For real time operation, the bottle neck for attacking with this technique comes in the identification and classification of the last meter measurement set within the wider pool i.e. the ability to identify which model the attack should be based upon. The proposed technique can also be practical for large systems which may contain a high number of measurements. In Figure \ref{largecomputationresults} we show the time to completion for the T-SNE and DBSCAN portions of the algorithm in the presence of very large random arrays (upto 10k points). It demonstrates the (expected) linear relationship from the T-SNE time complexity. Even when considering large data arrays with a large number of observations the time to perform the T-SNE/DBSCAN flow is relatively short. For example, using 10,000 observations for a 10,000 point system it took around 8.06 minutes using only an Intel Core i7-7820X CPU with 64GB of ram. We would expect that a highly motivated supranational attacker would have access to much more sophisticated hardware and would be able to execute such attacks even quicker.

\subsection{ Load Profile Bucketing}
As seen in \cite{Tian2019EnhancedGrids}, large load variations can make distinguishing changes from MTD in the system hard. Fundamentally, load variation can be used itself to hide MTD. Therefore we consider an extension to reduce a highly variate load system back to a steady loads style assumption under which the attacker can have more success. The attacker will use a combination of T-SNE and discrete bucketing to group load sets by their full system profile. The load variation within these individual buckets will be small and equivalent to a steady loads style assumption. The attacker can then run the attack over one of these buckets and it work as if the steady load assumption were in place.

\section{Physical Gaussian Watermarking with CUSUM}
While physical watermarking has not been applied in the power system space, the concept has been proposed in control systems such as in \cite{Mo2015PhysicalOutputs} where a watermark is added into a LQG-based control signals to drive detection. However, the papers in these areas aren't true 'physical' watermarks as they only change signal parameter dependencies and not the underlying physical plant itself. At the same  time, it should be noted that while MTD in the form of D-FACTS control to change system topology has been explored, the use of watermarks in combination with MTD has not been investigated and there is an opportunity to incorporate a true physical watermark into the system plant to enhance the system security. 


Previously, topology perturbation and transmission switching have been proposed as methods to drive detection of FDI attacks \cite{Morrow2012TopologyInjection}\cite{Tian2019EnhancedGrids}, 
These methods implement significant changes to line admittance as required by the change needed in residual (typically around 10-20\% for D-FACTS based changes), which may not only lead to interruption on system operation, but also provide opportunities for the data-driven attack to spot the existence of MTD and counter it. It is hence crucial that the deployment of MTD can remain hidden to the attacker. 

In this work, it is assumed that the SO will incorporate the capability of D-FACTS devices into the OPF model to optimize and select the lowest cost scenario as shown in \cite{Aghaei2012PlacementFlow}. MTD will then be applied around this point. As outlined in \cite{Lakshminarayana2020Cost-BenefitGrids}, there is a non-trivial cost incurred when applying conventional MTD. This cost comes in the form of non-optimal usage of power system assets. Where before D-FACTs are applied to minimise losses from reactive power they are now being used for MTD purposes away from this optimal point. As a result, the defender will wish to reduce the overall application of MTD.

In this context, this section proposes a novel method to achieve this by combining MTD with physical watermarking, which makes the MTD itself indistinguishable from the noise profile of the system, and monitoring sequential errors for long-run trends by using cumulative summed monitoring (CUSUM). CUSUM is a sequential analysis technique which monitors for change detection over a number of measurements. Samples taken from the process are assigned a weighting and summed to monitor change detection. In this case, we will monitor the measured residual $r$ under MTD defined by

\begin{equation} \label{residualhat}
          CEM_{t} =\sum_{j=1}^tr_j-T. 
\end{equation}
where $CEM$ is the decision statistic, $T$ is the target value of residual dictated by monitoring the statistic under normal conditions and $t$ is the number of periods in a measurement set, with upper and lower control limits $CEM^+_t$ and $CEM_t^-$. As $r$ is an absolute value, the lower bound $CEM_t^-$ will be 0. $CEM^+_t$ can be selected based on engineering judgement from prior observations. Usually the upper bound can be defined in terms of the residual variance and mean value under no attack:

\begin{equation} \label{residualhat}
          CEM^+_t = \bar{r}+B\sigma_r. 
\end{equation}
where $B$ is defined by the user based on previous observations and minimising type 2 error. 

The proposed defense strategy introduces these minor errors by using D-FACTS devices to alter the line admittance by a vector $\textbf{w}$. The size of admittance changes applied to each line is based on the output from a pseudo random number generator (PRNG), the seed value of which is only known by the network operator. This can be achieved with existing technology via a Unified Power Flow Controller (UPFC) 
in combination with a processing unit. 
The watermark may be applied selectively such that 

\begin{equation}
    w_m \in  \{0,\mathcal{N}(0,\,p)\}.
    \label{zattack}
\end{equation}
where $p$ is the max change applied to the branch admittance.

The resulting power flow profile under physical watermarking will be equal to

\begin{equation}
    \textbf{z}_w = (\textbf{H}+\textbf{w})\textbf{x} + \textbf{e}.
    \label{zattack}
\end{equation}
where $\textbf{w}$ represents the vector of admittance changes applied to branches and is known to the SO. 

\begin{figure}[t]
\centering
\includegraphics[width=2.8in]{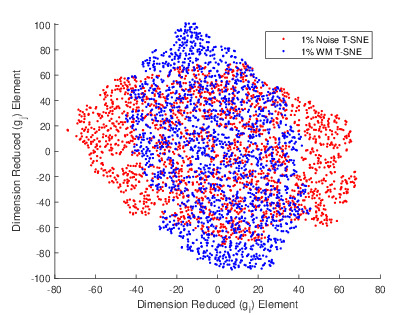}

\caption{Power flow profile observations of 1\% Gaussian watermark MTD applied to 14 lines intermittently under T-SNE dimensionality reduction. The X and Y axis values are non-dimensional probabilistic reductions ($g_i$ \& $g_j$ from equation \ref{dimensionreduc}). The system is reduced from a 34 dimension meter IEEE 14-bus system.}
\label{gaussianvstp2}
\end{figure}

The impact of applying a Gaussian style watermark in physical system parameters is shown in Figure \ref{gaussianvstp2}. Compared with direct binary perturbation, the proposed MTD show similar profile as underline noise and make it extremely hard for clustering algorithm to identify the existence of MTD or to counter it.

The key advantages of the proposed defense mechanism can be summarised as below, which will be validated in the next session:

\begin{enumerate}
    \item As the proposed MTD is on magnitude with the noise levels, the change in power flow observations resulting from the MTD becomes difficult to be identified. Therefore, MTD stays stealthy to the attacker.  
    \item Due to the stealthiness of the proposed MTD, it significantly increases the chance of the detection of FDI attack and is specifically resilient to intelligent attack types such as the proposed DBSCAN blind-ICA attack.
    \item The significantly-reduced magnitude of topology changes lead to less interruptions on the system stability and economic operation.
\end{enumerate}


\section{Results and Analysis}

This section assesses the performance of the proposed intelligent blind FDI in the presence of different forms of MTD on the standard IEEE 14-Bus and IEEE 118-bus test systems \cite{ChristiePowerArchive}. All simulations were implemented using the MATPOWER toolbox in MATLAB \cite{Zimmerman2011MATPOWER:Education} and  performed using Intel Core i7-7820X CPU with 64GB of ram running on a Windows 10 system. In the graph legends, TP refers to topology perturbation (MTD via D-FACTs perturbation) and RS refers to switching MTD via circuit breaker control. 




\subsection{Model Assumptions}
The priority of this section is to capture the change in detection between the blind FDI technique and the proposed intelligent attack under different types of MTD. Some assumptions have been made across all simulations:

\begin{itemize}
    \item Uncoloured Gaussian noise error of 1\% noise-to-signal was added to meter values as error $\textbf{e}$ as seen previously in \cite{Khanna2018AIGrid}.
    \item A steady load assumption is made with load variation of around 0.1\% for initial simulations as seen in as \cite{Tian2019EnhancedGrids}. Additional case studies were performed with multiple load profiles. 
    \item A minimum number of observations of 250 is assumed initially which rises to 1000 sequentially over the course of the simulation.
\end{itemize}


\begin{figure}[t]
\centering
\includegraphics[width=2.8in]{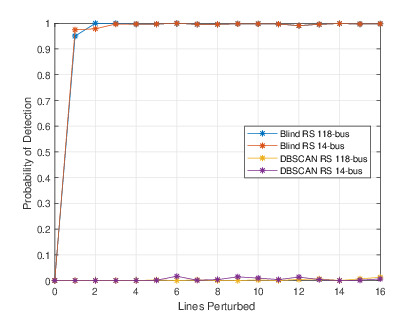}

\caption{Probability of detection of blind FDI attack and the new attack under transmission switching for IEEE 14-bus and 118-bus systems under 99\% confidence interval. Lines are not perturbed simultaneously.}
\label{TSresults}
\end{figure}

\begin{figure}[t]
\centering
\includegraphics[width=2.8in]{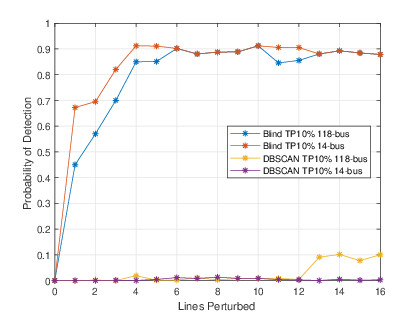}

\caption{Probability of detection of blind FDI attack and the new attack under admittance perturbation for IEEE 14-bus and 118-bus systems under 99\% confidence interval. Lines are not perturbed simultaneously.  }
\label{TPresults}
\end{figure}




\subsection{Line Applications of MTD}

In this paper, MTD is applied at the branch level in a fixed order shown in table \ref{linelist} to inductance values, based on a \% of the branch inductance. This order is consistent between MTD type to ensure a fair comparison between the MTD performance. Number of lines perturbed NLP refers to the number of adjusted lines within a given scenario. Line adjustments are not applied simultaneously and therefore a simulation will have $NLP+1$ potential underlying topologies that a successful attack will need to model for. The NLP perturbation list is additive and the topologies within them randomly selected from within this list. 

\begin{table}[]
\centering
\begin{tabular}{|l|l|l|l|l|}
\hline
\multicolumn{1}{|c|}{\textbf{Bus 1}} & \multicolumn{1}{c|}{\textbf{Bus 2}} & \multicolumn{1}{c|}{\textbf{R}} & \multicolumn{1}{c|}{\textbf{X}} & \multicolumn{1}{c|}{\textbf{NLP}} \\ \hline
1                                    & 2                                   & 0.01938                         & 0.05917                         & 1                                 \\ \hline
1                                    & 5                                   & 0.05403                         & 0.22304                         & 2                                 \\ \hline
2                                    & 3                                   & 0.04699                         & 0.19797                         & 3                                 \\ \hline
2                                    & 4                                   & 0.05811                         & 0.17632                         & 4                                 \\ \hline
2                                    & 5                                   & 0.05695                         & 0.17388                         & 5                                 \\ \hline
3                                    & 4                                   & 0.06701                         & 0.17103                         & 6                                 \\ \hline
4                                    & 5                                   & 0.01335                         & 0.04211                         & 7                                 \\ \hline
4                                    & 7                                   & 0                               & 0.20912                         & 8                                 \\ \hline
4                                    & 9                                   & 0                               & 0.55618                         & 9                                 \\ \hline
5                                    & 6                                   & 0                               & 0.25202                         & 10                                \\ \hline
6                                    & 11                                  & 0.09498                         & 0.1989                          & 11                                \\ \hline
6                                    & 12                                  & 0.12291                         & 0.25581                         & 12                                \\ \hline
6                                    & 13                                  & 0.06615                         & 0.13027                         & 13                                \\ \hline
9                                    & 10                                  & 0.03181                         & 0.0845                          & 14                                \\ \hline
9                                    & 14                                  & 0.12711                         & 0.27038                         & 15                                \\ \hline
10                                   & 11                                  & 0.08205                         & 0.19207                         & 16                                \\ \hline
\end{tabular}
\caption{Order of Number of lines perturbed (NLP) applied.}
\label{linelist}
\end{table}

\subsection{Transmission Switching}
The first form of MTD we trial is direct use of system circuit breakers to create new topologies (transmission switching). In this case, lines are switched into and out of operation to change the underlying topology incidence matrix. This creates significant changes in the overall power measurement matrix.
Figure \ref{TSresults} shows the impact of transmission line switching on the blind FDI attack and DBSCAN attack for the 14 bus and 118 bus cases. For the standard blind FDI attack, transmission switching is highly effective at introducing residual errors and driving alarms. With a single line switching the detection is 100\% for the standard blind FDI attack. However, these large changes in the system flows make it easy for an attacker to identify the MTD. Compared with the standard attack, the DBSCAN attack out-performs the standard blind FDI whenever MTD is used. Detection remained low (less than 1\%) with up to 15 lines being switched in/out across the network at different times. Even with 16 possible topologies in use the detection remained under 3\%. Transmission switching is unlikely to be used for the sole purpose of attack detection due to the significant impact on the system operability.

\subsection{Admittance Perturbation}

Admittance perturbation is the most commonly proposed method of MTD for power systems in the current literature. This sub-section implements an admittance perturbation defense against the typical blind FDI attack and the proposed DBSCAN version. A quantity equal to 10\% branch admittance is injected into the lines given by the order in \ref{linelist}. As discussed, branch admittance are applied independently and the number of underlying topologies will be equal to $NLP+1$. When the inductance is injected, the system operator is expecting to see the change in admittance reflected in the resulting power flows. If the attacker is unaware and does not reflect the new admittance in their attacking vector, the residual will increase significantly and BDD will be triggered. The results of admittance perturbation on detection of the standard and DBSCAN blind FDI attack are shown in Figure \ref{TPresults}. System models with branch admittance perturbations of 10\% were implemented. The standard blind FDI attack performs poorly against this form of MTD. For a single line at 10\% perturbation a detection level of over 95\% is achieved. The detection rates for the DBSCAN informed attack were consistently low. This is due to the distinctive clusters of power flows emerging under the steady loads assumption. There is a small spike which appears around 12 lines perturbed. This is likely due to the increasing number of lines perturbed in the system likely causing a misclustering in the underlying data-set or depriving a cluster of enough data points for a decent model.

\subsection{Physical Gaussian Watermarking with Cumulative Errors}

A novel MTD implementation is trialled here. In the same order and manner as the transmission switching and admittance perturbation sections we apply a Gaussian style physical watermark as defence. Inductance change of 1\% to the system varied over a random distribution. Only one line change is applied at a time to keep it consistent with the other forms of MTD.  The admittance profile is varied using a PRNG with a profile equivalent to the underlying noise of the system. As we have used a 1\% noise for our simulations the $p$ value is set equal to 1\% to ensure that this profile is not visible to the attacker. This is combined with cumulative summed error monitoring watching for sustained increased errors over 10 measurements with a cumulative limit based on 2 standard deviations above the average CUSUM measurement error summation under normal conditions.

\begin{figure}[t]
\centering
\includegraphics[width=2.8in]{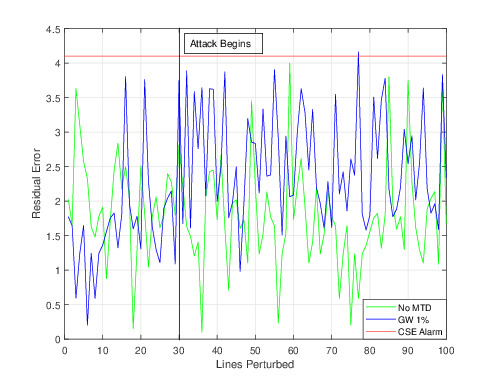}

\caption{Conventional CSE Residual error for run numbers on 14-bus system with the Gaussian Watermark applied to 14 lines. Bus angle change of 20 degrees attempted across the system by the FDI attack. }
\label{cusumresidual}
\end{figure}

\begin{figure}[t]
\centering
\includegraphics[width=2.8in]{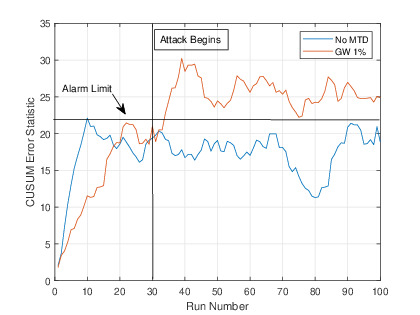}

\caption{CUSUM rolling summations for run 14-bus system with the Gaussian Watermark applied to 14 lines. Bus angle change of 20 degrees attempted across the system by the FDI attacker. }
\label{cusumaverage}
\end{figure}

Figures  \ref{cusumresidual} \& \ref{cusumaverage} illustrate the implementation of the Gaussian Watermark with assumption that FDI attack starts from time instance 30. Figure \ref{cusumresidual} shows the traditional CSE residual error resulting from a FDI attack in presence of the Gaussian Watermark. It is clear that the small system changes can not directly drive the detection of FDI attack in conventional residual-based BDD. Monitoring for the average of last 10 measurements allows the system operator to identify long term trends in the data, which in this case are caused by small but sustained gross errors introduced from the FDI attack. In Figure \ref{cusumaverage} the CUSUM method of detection is applied based on the last 10 measurements. Initially, we do not implement any attack for the first 30 runs of the system and we note residual CUSUM averages in line the normal value. At run 30 we introduce the attack vector. From inspection, it is much clearer that the system is under attack and a alarm is raised after 4 consecutive measurements.

\begin{figure}[t]
\centering
\includegraphics[width=2.8in]{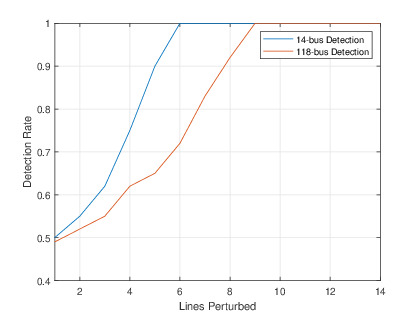}

\caption{DBSCAN detection results under proposed Gaussian watermark with cumulative errors over 10 measurements. 14-bus and 118-bus systems simulated with baseline 10 measurement average as detection trigger.}
\label{gaussianresults1}
\end{figure}

\begin{figure}[t]
\centering
\includegraphics[width=2.8in]{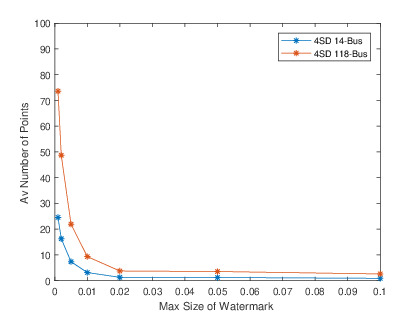}

\caption{Average number of points required to break a 4 standard deviation upper limit for increasing size of watermark applied to a single line.}
\label{avnumberpoints}
\end{figure}

As shown in Figure \ref{gaussianresults1}, under the DBSCAN blind FDI attack, the CUSUM Gaussian watermark shows significant improvements. As additional lines are added these detection rates are close to 100\% compared with under 10\% for standard admittance perturbation. The is due to the difficulty DBSCAN algorithm has in identifying clusters for MTD on magnitude and identical to noise profile of the system. Type-II error based on 2 standard deviation moves from the CUSUM average appears to give around 3\% type-II error for this kind of measurement approach across 1000 measurements. As seen in \ref{gaussianresults1} this cumulative approach also requires multiple measurements which potentially could lead to the attacker having additional time to attack before being caught. Therefore, there is a trade off between the speed to spot attacks and the magnitude of the added watermark. Figure \ref{avnumberpoints} illustrates this for the 118-bus and 14-bus systems where for a lower level of added watermark, a larger number of measurement points are needed to break the threshold.

\subsection{Load Variance Impact}
\begin{figure}[t]
\centering
\includegraphics[width=2.8in]{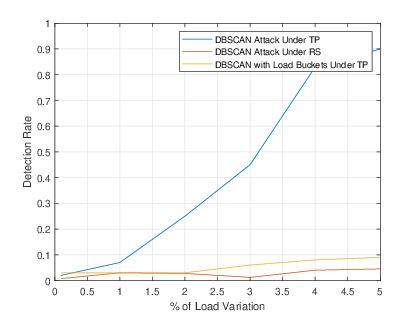}

\caption{Detection of DBSCAN method with 10 lines perturbed with increasing load variance. Also featured is the DBSCAN with load profile reduction analysis with load variation effectively reduced using 10 load buckets.}
\label{loadvar}
\end{figure}

\begin{figure}[t]
\centering
\includegraphics[width=2.8in]{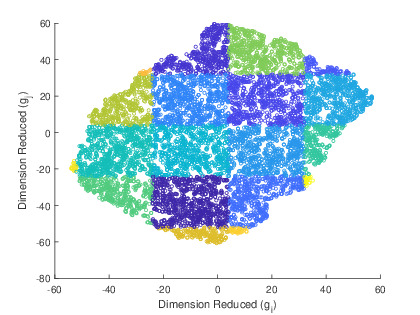}

\caption{Real power load profile values reduced by T-SNE. Variation of 10\% shown. Different colours represent different proposed load buckets. 10k measurements.}
\label{loadbins}
\end{figure}

\begin{figure}[t]
\centering
\includegraphics[width=2.8in]{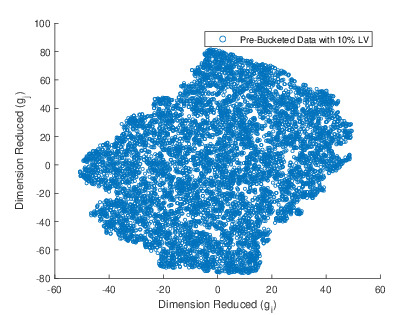}

\caption{Power Observations for a 10 lines perturbed system using D-FACTs of 10\% under load variance of 10\% shown. This is prior to bucketing of data by load profile.}
\label{loadbins2}
\end{figure}

\begin{figure}[t]
\centering
\includegraphics[width=2.8in]{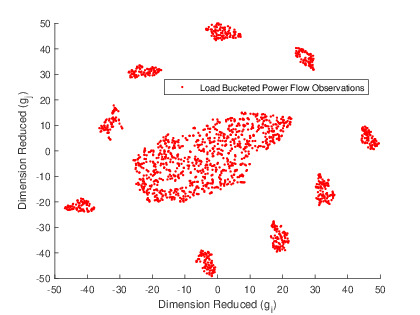}

\caption{Post-bucketed power flow data with T-SNE applied for a 10 lines perturbed system using D-FACTs of 10\%. Original load variance of 10\% was used.}
\label{loadbins3}
\end{figure}

 In previous case studies, the simulations have been performed under steady load assumptions \cite{Tian2019EnhancedGrids}. This session investigates the impact of large load variation on the performance of the DBSCAN attack. As shown in Figure \ref{loadvar}, large load variations reduce the effectiveness of the DBSCAN under pressure from topology perturbation due to the increasing challenge to cluster the topology changes under high varying load. To circumvent this challenge, we separate differing load values into buckets based on the system profiles reduced via T-SNE. Load values are observed directly, dimension reducing them via T-SNE and assigning them to bins of similar values. In Figure \ref{loadbins} the profile of the loads themselves are observed, dimension reduced and bucketed. Within each load groups, measurement observations can be used to obtain the clear MTD groups to develop the attack model. Load bucketing reduces the effective load variation back down to a steady loads style scenario. Figure \ref{loadbins2} and Figure \ref{loadbins3} shows the results of high load variation on the system under MTD with and without load bucketing applied. It is clear that under load bucketing, the distinct groups of measurement observations becomes clearer as a result of MTD. Applying this bucketing reduces the effective variance at the power flow significantly from 10\% to under 1\% with around 25 buckets for a 14-bus system. This effectively replicates the steady loads assumption even in the case of a more variate system. As the system variance becomes larger, additional bucket can be added to accommodate the larger variance of the system. The effect of this can be seen in Figure \ref{loadvar} with lowered detection for the DBSCAN attack when implemented against topology perturbation style defence. Bucketing in this manner will require a large amount of data, however based on the frequency of measurement at around 2-5s \cite{BudkaComputerReal} and the lengthy attack development phase \cite{Liang2017TheAttacks}, such data requirement can be easily satisfied. Blind attacks itself will always require larger past data requirement than full knowledge attacks as they need to build a model, unlike the full knowledge attacks which already possess the model. 

\subsection{Blind AC Replay Attack}
\begin{figure}[t]
\centering
\includegraphics[width=2.8in]{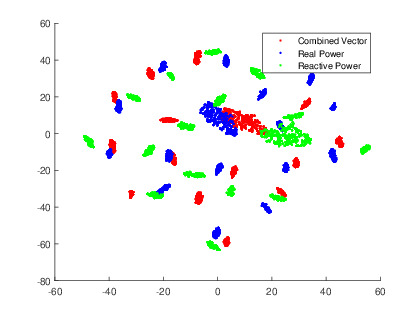}

\caption{Observations of 1\% MTD applied to AC system up to 16 lines intermittently. Data cuts of real power, reactive power and a combined vector incorporating both are compared. 1\% Gaussian noise assumed.}
\label{acobvs}
\end{figure}

\begin{figure}[t]
\centering
\includegraphics[width=2.8in]{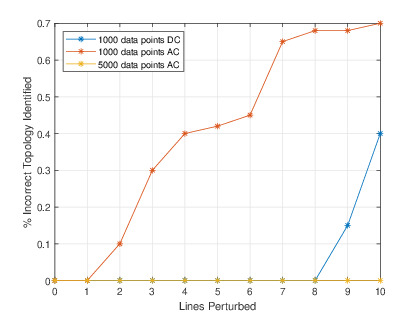}

\caption{AC System probability of wrong cluster identified for in presence of D_FACTs MTD with increasing lines perturbed to 14-bus system}
\label{acobvs2}
\end{figure}

We have implemented our clustering approach with a blind replay style attack against an AC state estimation. Under this attack the attacker attempts to inject a previously observed vector. The attacker is competing with MTD and wants to select the replay vector from a pool of values only containing those using the same topology configuration. In Figure \ref{acobvs} we can see that the distinctive cluster relationship exists within the AC model as shown previously for DC. Figure \ref{acobvs2} demonstrates that the proposed pre-clustering algorithm preforms well in AC state estimation provided a large number of samples received. The non-linearity in the AC model significantly reduce the correction rate of clustering but increasing the number of observations allows good performance for the AC model.

\section{Conclusions \& Further Work}
This paper, for the first time, investigate how unsupervised learning and dimensionality reduction can be applied in blind FDI attacks to exploit the venerability of current forms of MTD.  By incorporating a combination of T-SNE dimensionality reduction and the DBSCAN clustering algorithm, power flow observations can be clustered into their relative topology profiles and the mixing matrix for the blind FDI attack can be calculated using only data under the same network topology. This technique is shown to be effective against admittance perturbation and transmission switching techniques. A novel defense strategy against this new type attack is proposed through combining MTD with physical watermarking to add  indistinguishable Gaussian style physical watermark into the plant topology and  monitoring the sequential  errors  for long-run trends by using CUSUM. This technique is demonstrated to be effective at both inhibiting the attacker's ability to predict topological changes from visible power flows and reducing the overall impact on system operation by reducing the level of topology changes. 

 Further work on this topic entails enhancing the blind FDI model to model for other scenarios i.e. subset attacks, optimal design of physical watermarking scheme and analysing the effects of MTD on topological discovery techniques.

\ifCLASSOPTIONcaptionsoff
  \newpage
\fi

\bibliographystyle{IEEEtran.bst}
\bibliography{bare_jrnl.bbl}

\begin{thebibliography}{10}
\providecommand{\url}[1]{#1}
\csname url@samestyle\endcsname
\providecommand{\newblock}{\relax}
\providecommand{\bibinfo}[2]{#2}
\providecommand{\BIBentrySTDinterwordspacing}{\spaceskip=0pt\relax}
\providecommand{\BIBentryALTinterwordstretchfactor}{4}
\providecommand{\BIBentryALTinterwordspacing}{\spaceskip=\fontdimen2\font plus
\BIBentryALTinterwordstretchfactor\fontdimen3\font minus
  \fontdimen4\font\relax}
\providecommand{\BIBforeignlanguage}[2]{{%
\expandafter\ifx\csname l@#1\endcsname\relax
\typeout{** WARNING: IEEEtran.bst: No hyphenation pattern has been}%
\typeout{** loaded for the language `#1'. Using the pattern for}%
\typeout{** the default language instead.}%
\else
\language=\csname l@#1\endcsname
\fi
#2}}
\providecommand{\BIBdecl}{\relax}
\BIBdecl

\bibitem{Liang2017TheAttacks}
G.~Liang, S.~R. Weller, J.~Zhao, F.~Luo, and Z.~Y. Dong, ``{The 2015 Ukraine
  Blackout: Implications for False Data Injection Attacks},'' \emph{IEEE
  Transactions on Power Systems}, 2017.

\bibitem{Liu2011FalseGrids}
Y.~Liu, P.~Ning, and M.~K. Reiter, ``{False data injection attacks against
  state estimation in electric power grids},'' \emph{ACM Transactions on
  Information and System Security}, 2011.

\bibitem{Liu2016MaskingAttacks}
X.~Liu, Z.~Li, X.~Liu, and Z.~Li, ``{Masking Transmission Line Outages via
  False Data Injection Attacks},'' \emph{IEEE Transactions on Information
  Forensics and Security}, vol.~11, no.~7, pp. 1592--1602, 7 2016.

\bibitem{Deng2017FalseSurvey}
R.~Deng, G.~Xiao, R.~Lu, H.~Liang, and A.~V. Vasilakos, ``{False data injection
  on state estimation in power systems-attacks, impacts, and defense: A
  survey},'' \emph{IEEE Transactions on Industrial Informatics}, 2017.

\bibitem{Liang2017ASystems}
G.~Liang, J.~Zhao, F.~Luo, S.~R. Weller, and Z.~Y. Dong, ``{A Review of False
  Data Injection Attacks Against Modern Power Systems},'' \emph{IEEE
  Transactions on Smart Grid}, 2017.

\bibitem{Rahman2012FalseGrids}
M.~A. Rahman and H.~Mohsenian-Rad, ``{False data injection attacks with
  incomplete information against smart power grids},'' in \emph{GLOBECOM - IEEE
  Global Telecommunications Conference}, 2012.

\bibitem{Esmalifalak2011StealthGrid}
M.~Esmalifalak, H.~Nguyen, R.~Zheng, and Z.~Han, ``{Stealth false data
  injection using independent component analysis in smart grid},'' in
  \emph{2011 IEEE International Conference on Smart Grid Communications,
  SmartGridComm 2011}, 2011.

\bibitem{Deng2019FalseGrid}
R.~Deng and H.~Liang, ``{False Data Injection Attacks with Limited Susceptance
  Information and New Countermeasures in Smart Grid},'' \emph{IEEE Transactions
  on Industrial Informatics}, vol.~15, no.~3, pp. 1619--1628, 3 2019.

\bibitem{Kurt2018Real-TimeGridb}
M.~N. Kurt, Y.~Yilmaz, and X.~Wang, ``{Real-Time Detection of Hybrid and
  Stealthy Cyber-Attacks in Smart Grid},'' \emph{IEEE Transactions on
  Information Forensics and Security}, vol.~14, no.~2, pp. 498--513, 2 2018.

\bibitem{Wang2017AGrids}
Y.~Wang, M.~M. Amin, J.~Fu, and H.~B. Moussa, ``{A novel data analytical
  approach for false data injection cyber-physical attack mitigation in smart
  grids},'' \emph{IEEE Access}, vol.~5, pp. 26\,022--26\,033, 11 2017.

\bibitem{Ahmed2019UnsupervisedForest}
S.~Ahmed, Y.~Lee, S.~H. Hyun, and I.~Koo, ``{Unsupervised Machine
  Learning-Based Detection of Covert Data Integrity Assault in Smart Grid
  Networks Utilizing Isolation Forest},'' \emph{IEEE Transactions on
  Information Forensics and Security}, vol.~14, no.~10, pp. 2765--2777, 10
  2019.

\bibitem{Wen2019AnGrids}
F.~Wen and W.~Liu, ``{An Efficient Data-Driven False Data Injection Attack in
  Smart Grids},'' in \emph{International Conference on Digital Signal
  Processing, DSP}, vol. 2018-November.\hskip 1em plus 0.5em minus 0.4em\relax
  Institute of Electrical and Electronics Engineers Inc., 1 2019.

\bibitem{Kim2015SubspaceApproach}
J.~Kim, L.~Tong, and R.~J. Thomas, ``{Subspace methods for data attack on state
  estimation: A data driven approach},'' \emph{IEEE Transactions on Signal
  Processing}, vol.~63, no.~5, pp. 1102--1114, 3 2015.

\bibitem{Hao2015SparseGrids}
J.~Hao, R.~J. Piechocki, D.~Kaleshi, W.~H. Chin, and Z.~Fan, ``{Sparse
  Malicious False Data Injection Attacks and Defense Mechanisms in Smart
  Grids},'' \emph{IEEE Transactions on Industrial Informatics}, vol.~11, no.~5,
  pp. 1198--1209, 10 2015.

\bibitem{Zhang2018CanSystems}
J.~Zhang, Z.~Chu, L.~Sankar, and O.~Kosut, ``{Can attackers with limited
  information exploit historical data to mount successful false data injection
  attacks on power systems?}'' \emph{IEEE Transactions on Power Systems}, 2018.

\bibitem{Wang2015EffectsNetworks}
S.~Wang, W.~Ren, and U.~M. Al-Saggaf, ``{Effects of Switching Network
  Topologies on Stealthy False Data Injection Attacks Against State Estimation
  in Power Networks},'' \emph{IEEE Systems Journal}, vol.~11, no.~4, pp.
  2640--2651, 11 2015.

\bibitem{Morrow2012TopologyInjection}
K.~L. Morrow, E.~Heine, K.~M. Rogers, R.~B. Bobba, and T.~J. Overbye,
  ``{Topology Perturbation for Detecting Malicious Data Injection},'' in
  \emph{2012 45th Hawaii International Conference on System Sciences}, 2012.

\bibitem{Liu2017ReactanceEstimation}
C.~Liu, M.~Zhou, J.~Wu, C.~Long, A.~Farraj, E.~Hammad, and D.~Kundur,
  ``{Reactance Perturbation for Enhancing Detection of FDI Attacks in Power
  System State Estimation},'' in \emph{2017 IEEE Global Conference on Signal
  and Information Processing (GlobalSIP)}, 2017.

\bibitem{Zhang2019AnalysisGrid}
Z.~Zhang, R.~Deng, D.~Yau, P.~Cheng, and J.~Chen, ``{Analysis of Moving Target
  Defense Against False Data Injection Attacks on Power Grid},'' \emph{IEEE
  Transactions on Information Forensics and Security}, 2019.

\bibitem{Li2019OnDevices}
B.~Li, G.~Xiao, R.~Lu, R.~Deng, and H.~Bao, ``{On Feasibility and Limitations
  of Detecting False Data Injection Attacks on Power Grid State Estimation
  Using D-FACTS Devices},'' \emph{IEEE Transactions on Industrial Informatics},
  pp. 1--1, 6 2019.

\bibitem{Tian2019EnhancedGrids}
J.~Tian, R.~Tan, X.~Guan, and T.~Liu, ``{Enhanced hidden moving target defense
  in smart grids},'' \emph{IEEE Transactions on Smart Grid}, vol.~10, no.~2,
  pp. 2208--2223, 3 2019.

\bibitem{Mo2015PhysicalOutputs}
Y.~Mo, S.~Weerakkody, and B.~Sinopoli, ``{Physical authentication of control
  systems: Designing watermarked control inputs to detect counterfeit sensor
  outputs},'' \emph{IEEE Control Systems}, vol.~35, no.~1, pp. 93--109, 2 2015.

\bibitem{Monticelli1999StateApproach}
A.~Monticelli, \emph{{State Estimation in Electric Power Systems: A Generalized
  approach}}, 1999.

\bibitem{Yu2015BlindGrid}
Z.~H. Yu and W.~L. Chin, ``{Blind False Data Injection Attack Using PCA
  Approximation Method in Smart Grid},'' \emph{IEEE Transactions on Smart
  Grid}, 2015.

\bibitem{Chin2018BlindCommunications}
W.~L. Chin, C.~H. Lee, and T.~Jiang, ``{Blind false data attacks against ac
  state estimation based on geometric approach in smart grid communications},''
  \emph{IEEE Transactions on Smart Grid}, vol.~9, no.~6, pp. 6298--6306, 11
  2018.

\bibitem{VanDerMaaten2008VisualizingT-SNE}
L.~Van Der~Maaten and G.~Hinton, ``{Visualizing Data using t-SNE},'' Tech.
  Rep., 2008.

\bibitem{ESilva2010MachineTesting}
D.~G. E~Silva, M.~Jino, and B.~T. De~Abreu, ``{Machine learning methods and
  asymmetric cost function to estimate execution effort of software testing},''
  in \emph{ICST 2010 - 3rd International Conference on Software Testing,
  Verification and Validation}, 2010, pp. 275--284.

\bibitem{Pezzotti2017ApproximatedAnalytics}
N.~Pezzotti, B.~P. Lelieveldt, L.~Van Der~Maaten, T.~H{\"{o}}llt, E.~Eisemann,
  and A.~Vilanova, ``{Approximated and user steerable tSNE for progressive
  visual analytics},'' \emph{IEEE Transactions on Visualization and Computer
  Graphics}, vol.~23, no.~7, pp. 1739--1752, 7 2017.

\bibitem{McInnesBenchmarkingAlgorithms}
\BIBentryALTinterwordspacing
L.~McInnes, J.~Healy, and S.~Astels, ``{Benchmarking Performance and Scaling of
  Python Clustering Algorithms}.'' [Online]. Available:
  \url{https://hdbscan.readthedocs.io/en/latest/}
\BIBentrySTDinterwordspacing

\bibitem{Ester1996ANoise}
\BIBentryALTinterwordspacing
M.~Ester, H.-P. Kriegel, J.~Sander, and X.~Xu, ``{A Density-Based Algorithm for
  Discovering Clusters in Large Spatial Databases with Noise},'' Tech. Rep.,
  1996. [Online]. Available: \url{www.aaai.org}
\BIBentrySTDinterwordspacing

\bibitem{Konstantinou2019AAttacks}
C.~Konstantinou and M.~Maniatakos, ``{A Data-Based Detection Method Against
  False Data Injection Attacks},'' \emph{IEEE Design and Test}, 2019.

\bibitem{Aghaei2012PlacementFlow}
J.~Aghaei, M.~Gitizadeh, and M.~Kaji, ``{Placement and operation strategy of
  FACTS devices using optimal continuous power flow},'' \emph{Scientia
  Iranica}, vol.~19, no.~6, pp. 1683--1690, 12 2012.

\bibitem{Lakshminarayana2020Cost-BenefitGrids}
\BIBentryALTinterwordspacing
S.~Lakshminarayana and D.~K.~Y. Yau, ``{Cost-Benefit Analysis of Moving-Target
  Defense in Power Grids},'' \emph{IEEE Transactions on Power Systems}, pp.
  1--1, 2020. [Online]. Available:
  \url{https://ieeexplore.ieee.org/document/9144465/}
\BIBentrySTDinterwordspacing

\bibitem{ChristiePowerArchive}
\BIBentryALTinterwordspacing
R.~Christie, ``{Power Systems Test Case Archive}.'' [Online]. Available:
  \url{http://labs.ece.uw.edu/pstca/}
\BIBentrySTDinterwordspacing

\bibitem{Zimmerman2011MATPOWER:Education}
R.~D. Zimmerman, C.~E. Murillo-S{\'{a}}nchez, and R.~J. Thomas, ``{MATPOWER:
  Steady-state operations, planning, and analysis tools for power systems
  research and education},'' \emph{IEEE Transactions on Power Systems}, 2011.

\bibitem{Khanna2018AIGrid}
K.~Khanna, B.~Panigrahi, and A.~Joshi, ``{AI based approach to identify
  compromised meters in data integrity attacks on smart grid},'' \emph{IET
  Generation, Transmission {\&} Distribution}, vol.~12, no.~5, pp. 1052 --
  1066, 2018.

\bibitem{BudkaComputerReal}
\BIBentryALTinterwordspacing
K.~C. Budka, J.~G. Deshpande, and M.~Thottan, ``{Computer Communications and
  Networks Communication Networks for Smart Grids Making Smart Grid Real},''
  Tech. Rep. [Online]. Available: \url{http://www.springer.com/series/4198}
\BIBentrySTDinterwordspacing

\end{thebibliography}

\begin{IEEEbiography}[{\includegraphics[width=1in,height=1.25in,clip,keepaspectratio]{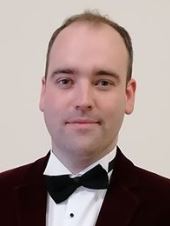}}]{Martin Higgins}
(S’19) received the BSc in Physics from Queen Mary, University of London, in 2011, and an MSc from Imperial College London, U.K., in 2012. Currently, he is pursuing a Ph.D. in Electrical Engineering at Imperial College London as part of the CDT in Smart Grids collaboration integrated MRES and PHD with the University of Strathclyde. His research interests lie in power systems cyber-security, false data injection attacks and moving target defence. 
\end{IEEEbiography}

\begin{IEEEbiography}[{\includegraphics[width=1in,height=1.25in,clip,keepaspectratio]{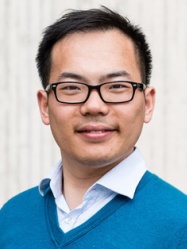}}]{Dr Fei Teng}
(M’15)  graduated with a BEng from Beihang University, China in 2009 and obtained Ph.D. from Imperial College London in 2015.  Currently,
he is a lecturer in the Department of Electrical and Electronic Engineering,
Imperial College London, U.K. His research focus on the efficient and resilient operation of future cyber-physical power system.
\end{IEEEbiography}

\begin{IEEEbiography}[{\includegraphics[width=1in,height=1.25in,clip,keepaspectratio]{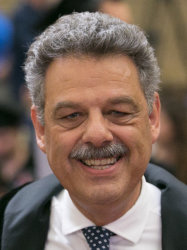}}]{Professor Thomas Parisini}
received the Ph.D. degree in Electronic Engineering and Computer Science in 1993 from the University of Genoa. He was with Politecnico di Milano and since 2010 he holds the Chair of Industrial Control and is Director of Research at Imperial College London. He is a Deputy Director of the KIOS Research and Innovation Centre of Excellence, University of Cyprus. Since 2001 he is also Danieli Endowed Chair of Automation Engineering with University of Trieste. In 2009-2012 he was Deputy Rector of University of Trieste. In 2018 he received an {\em Honorary Doctorate} from University of Aalborg, Denmark. He authored or co-authored more than 320 research papers in archival journals, book chapters, and international conference proceedings. His research interests include neural-network approximations for optimal control problems, distributed methods for cyber-attack detection and cyber-secure control of large-scale systems, fault diagnosis for nonlinear and distributed systems, nonlinear model predictive control systems and nonlinear estimation. He is a co-recipient of the IFAC Best Application Paper Prize of the Journal of Process Control, Elsevier, for the three-year period 2011-2013 and of the 2004 Outstanding Paper Award of the IEEE Trans. on Neural Networks. He is also a recipient of the 2007 IEEE Distinguished Member Award. In 2016, he was awarded as Principal Investigator at Imperial of the H2020 European Union flagship Teaming Project KIOS Research and Innovation Centre of Excellence led by University of Cyprus. In 2012, he was awarded an ABB Research Grant dealing with energy-autonomous sensor networks for self-monitoring industrial environments. Thomas Parisini currently serves as 2020 President-Elect of the IEEE Control Systems Society and has served as Vice-President for Publications Activities. During 2009-2016 he was the Editor-in-Chief of the IEEE Trans. on Control Systems Technology. Since 2017, he is Editor for Control Applications of Automatica and since 2018 he is the Editor in Chief of the European Journal of Control. He is also the Chair of the IFAC Technical Committee on Fault Detection, Supervision \& Safety of Technical Processes - SAFEPROCESS.  He was the Chair of the IEEE Control Systems Society Conference Editorial Board and a Distinguished Lecturer of the IEEE Control Systems Society. He was an elected member of the Board of Governors of the IEEE Control Systems Society and of the European Control Association (EUCA) and a member of the board of evaluators of the 7th Framework ICT Research Program of the European Union. Thomas Parisini is currently serving as an Associate Editor of the Int. J. of Control and served as Associate Editor of the IEEE Trans. on Automatic Control, of the IEEE Trans. on Neural Networks, of Automatica, and of the Int. J. of Robust and Nonlinear Control.  Among other activities, he was the Program Chair of the 2008 IEEE Conference on Decision and Control and General Co-Chair of the 2013 IEEE Conference on Decision and Control. Prof. Parisini is a Fellow of the IEEE and of the IFAC.
\end{IEEEbiography}

\end{document}